\documentclass{aa} 

\usepackage[varg]{txfonts}
\usepackage{natbib}
\usepackage{graphicx}
\usepackage{txfonts}
\newcommand{\oiii}{{[\ion{O}{iii}]}}

\newcommand{\coII}{\ion{CO}{(2-1)}}

\newcommand{\hii}{\ion{H}{ii}}

\newcommand{\ha}{H$\,\alpha$}

\newcommand{\cmcubic}{$\rm{cm^{-3}}$}

\newcommand{\kms}{{$\rm{km\,s^{-1}}$}}
\newcommand{\degree}{{$^{\circ}$}}
\newcommand{\ergs}{$\rm{erg\,s^{-1}}$}
\newcommand{\ergscmAA}{$\rm{erg\,s^{-1}\,cm^{-2}\,\si{\angstrom}}$}
\newcommand{\msun}{$\rm{M_{\odot}}$}

\newcommand{\jykms}{$\rm{Jy\,km\,s^{-1}}$}

\newcommand{\Kkmspc}{$\rm{K\,km\,s^{-1}\,pc^{-2}}$}
\newcommand{\msunKkmspc}{$\rm{M_{\odot}\,[K\,km\,s^{-1}\,pc^{-2}]^{-1}}$}
\usepackage{siunitx}
\DeclareSIUnit{\angstrom}{\textup{\AA}} 
\newcommand{\AAA}{\si{\angstrom}}

\newcommand{\vel}{{\rm{v}}}

\newcommand{\DL}{$D_{\rm{L}}$}
\newcommand{\sigmarms}{$\sigma_{rms}$}

\newcommand{\lagn}{$L_{\mathrm{AGN}}$}

\newcommand{\MBH}{$M_{\mathrm{BH}}$}
\newcommand{\Mstar}{$M_{\mathrm{*}}$}

\newcommand{\alphaCO}{$\alpha_{\rm{CO}}$}

\newcommand{\hst}{\textit{HST}}
\newcommand{\vla}{VLA}

\newcommand{\paperII}{Paper II}
\defcitealias{dallagnol+25_paperII}{Paper~II}
\defcitealias{dallagnol+25_paperI}{Paper~I}

\usepackage[dvipsnames]{xcolor}

\usepackage{hyperref} 
\hypersetup{
    colorlinks=true,
    citecolor=blue, 
    linkcolor=blue,
    filecolor=blue,      
    urlcolor=blue,
}

\usepackage{ulem} 
\usepackage{amsmath}	
\usepackage{amssymb}	

\usepackage{subcaption}  
\usepackage{newtxtext,newtxmath}

\begin{document}

   \title{NGC\,6860, Mrk\,915 and MCG\,-01-24-012}
   \subtitle{I. Spatial anti-correlation between cold molecular and ionized gas distributions in Seyfert galaxies}

    \titlerunning{NGC\,6860, Mrk\,915 and MCG\,-01-24-012. I. CO(2-1) vs. {\oiii} flux distribution}
    \authorrunning{Dall'Agnol de Oliveira et al.}

\author{Bruno Dall'Agnol de Oliveira\inst{\ref{inst1},\ref{inst2}}
   \and Thaisa Storchi-Bergmann\inst{\ref{inst2}}
   \and Neil Nagar\inst{\ref{inst3}}
   \and Santiago Garcia-Burillo\inst{\ref{inst4}}
   \and Rogemar A. Riffel\inst{\ref{inst5},\ref{inst5a}}
   \and Dominika Wylezalek\inst{\ref{inst1}}
   \and Pranav Kukreti\inst{\ref{inst1}}
   \and Venkatessh Ramakrishnan\inst{\ref{inst6}}
          }

   \institute{
Zentrum für Astronomie der Universität Heidelberg, Astronomisches Rechen-Institut, Mönchhofstr 12-14, D-69120 Heidelberg, Germany\label{inst1}\and
Departamento de Astronomia, Universidade Federal do Rio Grande do Sul, IF, CP 15051, 91501-970 Porto Alegre, RS, Brazil\label{inst2}\and
Astronomy Department, Universidad de Concepci\'on, Barrio Universitario S/N, Concepci\'on 4030000, Chile\label{inst3}\and
Observatorio Astronómico Nacional (OAN-IGN)-Observatorio de Madrid, Alfonso XII, 3, 28014 Madrid, Spain\label{inst4}\and
Departamento de F\'isica, CCNE, Universidade Federal de Santa Maria, 97105-900, Santa Maria, RS, Brazil\label{inst5}\and
Centro de Astrobiología (CAB), CSIC-INTA, Ctra. de Ajalvir km 4, Torrejón de Ardoz, 28850, Madrid, Spain\label{inst5a}\and
Finnish Centre for Astronomy with ESO, University of Turku, 20014 Turku, Finland\label{inst6}
   }

   \date{Accepted: 25 July 2025}

  \abstract
  {
We present a study of the cold molecular versus the ionized gas  distribution in 
three nearby Seyfert galaxies: NGC\,6860, Mrk\,915 and MCG\,-01-24-012.  
To constrain the cold molecular flux distribution at $\sim$\,0.5\,--\,0.8{\arcsec} ($\sim$\,150\,--\,400\,pc) scales, we used data from the CO(2–1) emission line, obtained with the Atacama Large Millimeter/submillimeter Array (ALMA).
For the ionized gas, we used \textit{Hubble Space Telescope} ({\hst}) narrow-band images, centered on the {\oiii}$\lambda\lambda$4959,5007 emission lines. 
Within the inner kiloparsec of the three galaxies, we observe gaps in the CO emission in regions co-spatial with the {\oiii} flux distribution, 
similarly to what has been recently observed in other active galaxies. From our original sample of 13 nearby AGN sources, 12 objects present the same trend. 
This indicates that CO molecules might be partially dissociated by AGN radiation or that there is a deficit of cold molecular gas on nuclear scales driven by ionized gas outflows/jets.
If so, this represents a form of AGN feedback that is not captured when only outflow kinematics properties, such as mass outflow rates, are considered. 
We also discuss that part of the molecular gas might still be present in hotter H$_2$ phases, as observed already in other objects. 
}


   \maketitle

\section{Introduction}\label{sec:intro}

Evaluating the impact of active galactic nuclei (AGN) on the interstellar medium (ISM) is crucial to understand their role in the evolution of galaxies. This interaction happens through the coupling of the energy released by AGN via radiation, winds and jets, as a product of the matter accreted to the central black hole \citep{heckman_best14}. 
However, to better gauge the AGN impact on the gas, we need to consider that different gas phases are present in the ISM, with all of them potentially being affected by the AGN to some degree  \citep{cicone+18}. 
In this work, we focus on two of them: the ionized and the cold molecular phase.

The {\oiii}$\lambda\lambda$4959,5007 emission lines doublet (hereafter, \oiii) has been widely used to gauge the AGN feedback on the ionized gas. 
The disturbed content manifests itself as large line widths (reaching values $>$\,10$^3$\,{\kms}) or broad emission line components
\citep[e.g][]{riffel+24,sun+17,karouzos+16,fischer+13,liu+13}. 
These forbidden lines require a high-energy photon source  -- like an AGN or young and massive stars -- and is an indicator of low-density gas tracer \citep{wylezalek+16}, with average gas temperatures and densities in AGN hosts of $T_{\rm{gas}}$\,$\sim$\,10$^4$\,K \citep[e.g.][]{revalski+21,riffelRA+21} and $n_{\rm{gas}}$\,$\sim$\,10$^3${\cmcubic} \citep[e.g.][]{davies+20,revalski+22}. 
For some AGN sources, narrow-band imaging observations show {\oiii} emission with bipolar morphologies (elongated in opposite/aligned directions) in regions photo-ionized by the AGN \citep[e.g.][]{storchi-bergmann+18,schmitt+03,wilson+93,storchi-bergmann+92, rodriguez-ardila+17}. 
This morphology is thought to result from the AGN ionization axis being more aligned with the plane of the sky, plus the dust in the galaxy disk and the torus blocking the light along the line-of-sight \citep{storchi-bergmann+18}. 
Later long-slit and integral field observations revealed the presence of outflows in these bipolar regions, with kinematically disturbed gas being detected up to an extent that is \,$\sim$\,20\,--\,50 per cent (on average)  of the extent of the total photoionized region  \citep[e.g.][]{dallagnol+21,travis+18}.

New stars are formed from the collapse of dense molecular clouds with $T_{\rm{gas}}$\,<\,100\,K and $n_{\rm{gas}}$\,>\,10$^{3}${\cmcubic}  \citep{saintonge+22, bolatto+13}.
These dense clouds are a subset of the total cold molecular gas content -- dominated by H$_2$ molecules -- that condensed from the atomic gas present in the galaxies' disks  \citep{schinnerer_leroy24}. 
To trace the total cold molecular mass and its spatial distribution, we can use emission lines from CO molecules, like the CO(2-1) transition (230.538\,GHz rest frequency).
In the last decade, these CO lines have been successfully used to identify AGN-driven disturbances  -- from jets and/or winds -- in the molecular phase  \citep[e.g.][]{dallagnol+23,ramos-almeida+22,garcia-bernete+21,slater+19,finlez+18}.

By comparing the flux distribution of both gas phases, we can look for signs of how each gas phase interacts with the AGN released energy. 
For this, we need to resolve the gas on scales of 100\,pc ($\sim$\,0.5{\arcsec}, for nearby sources). This can be achieved with observation of the CO lines taken with the \textit{Atacama Large Millimeter/submillimeter Array} (ALMA) observatory, and \textit{Hubble Space Telescope} ({\hst}) narrow-band imaging for the {\oiii} lines.
In this work, we gathered such data for three nearby Seyfert galaxies: \object{NGC\,6860}, \object{Mrk\,915} and \object{MCG\,-01-24-012}. The study is divided in two parts.

In this paper, we investigate the relationship between the cold molecular and ionized gas flux distributions. 
The modeling of the CO kinematics will be done in a companion paper \citet[hereafter \paperII]{dallagnol+25_paperII}. 
In Sect.\,\ref{sec:sample}, 
we describe the sample, while the observations and reduction of the {ALMA} data are discussed in Sect.\,\ref{sec:obs}, with some details added in Appendices\,\ref{ap:rgb} and \ref{ap:oiii-continuum}. 
The analysis and results are detailed in Sects.\,\ref{sec:analysis} and \ref{sec:results}, leaving the discussions and conclusions to Sects.\,\ref{sec:Discussion} and \ref{sec:conclusions}.

\begin{table*}
	\centering
    \caption{
    (1) Galaxy name; 
    (2) RA and DEC coordinates of the ALMA millimeter continuum peak (in the ICRS frame); 
    (3) Redshift, corresponding to the systemic velocity (in the LSRK frame); 
    (4) Luminosity distance; 
    (5) Angular scale; 
    (6) Seyfert type; 
    (7) AGN total luminosity; 
    (8) Stellar mass;  
    (9) Black Hole mass; 
    The range of values in {\MBH} and {\lagn} arises from the different measurements that we collected from the literature, with the exceptions calculated by us described below in the notes. In particular for {\lagn}, the ranges are partially due to an observed intrinsic X-Ray variability.
    } 
\begin{tabular}{lccccccccr}
\hline
Name &                  RA, DEC &       z &            \DL &                  Scale &    Type &             log({\Mstar}) &               log({\MBH}) &                log({\lagn}) \\
     &    hh:mm:ss.ss dd:mm:ss.ss            &         & $\mathrm{Mpc}$ & $\mathrm{pc/\arcsec}$ &         & $\mathrm{M_{\odot}}$ & $\mathrm{M_{\odot}}$ & $\mathrm{erg\,s^{-1}}$ \\
(1) & (2) & (3) & (4) & (5) & (6) & (7) & (8) & (9) \\
\hline
 NGC\,6860 & 20:08:46.89 -61:05:59.77 & 0.01477 &           64.0 &                  301 &     1.5$^{a}$ &     10.3$^d$        &     7.3\,--\,8.3$^e$ &  43.6\,--\,43.8$^{g,h}$ \\
 Mrk\,915 & 22:36:46.50 -12:32:42.80 & 0.02415 &            105 &                  487 & 1.5/1.9$^{b}$ &            10.0$^d$ &     7.3\,--\,8.4$^e$ &  44.1\,--\,44.3$^{g,h}$ \\
 MCG\,-01-24-012 & 09:20:46.26 -08:03:21.97 & 0.01972 &           85.7 &                  400 &   1.9/2$^{c}$ &   9.64$^d$  &      7.2$\pm$0.3$^f$ & 44.3\,--\,44.8$^{i,j}$ \\
\hline
\end{tabular}
{\raggedright
\\
$^{a}$: \citet{lipari+93}, with detected variability on the {\ha} broad line region (BRL) shape and UV continuum between 1989 and 1991.\par
$^{b}$: \citet{goodrich1995,bennert+06,trippe+10}. \par
$^{c}$: \citet{veron-cetty_veron06,onori+17,la_franca+15}.\par
$^{d}$: \citet{hernandez-yevenes+24}, from \textit{Wide-field Infrared Survey Explorer} (\textit{WISE}) photometry, using W1 and the W1-W2 color for M/L corrections. \par
$^{e}$: \citet{bennert+06}, range from different methods, including the \MBH-$\sigma_*$ relation and from L$_{\rm{5100\,}\AAA}$. \par 
$^{f}$: \citet{la_franca+15}, from Pa$\beta$ BRL line. \par
$^{g}$: \citet{bennert+06}, from L$_{\rm{5100\,\AAA}}$. \par
$^{h}$: Obtained by us using the \citet[0.37 dex]{duras+20} correction with L$_{\rm{2-10\,KeV}}$ values from \citet[Mrk\,915]{ballo+17} and \citet[NGC\,6860]{winter+09}. \par
$^{i}$: \citet{middei+21}, from L$_{\rm{2-10\,KeV}}$. \par 
$^{j}$: \citet{la_franca+15}, from L$_{\rm{14-195\,KeV}}$. \par
}
\label{tab:sample}
\end{table*}

\section{Sample}\label{sec:sample}

\begin{figure*}
     \centering
    \includegraphics[width=0.383\textwidth]{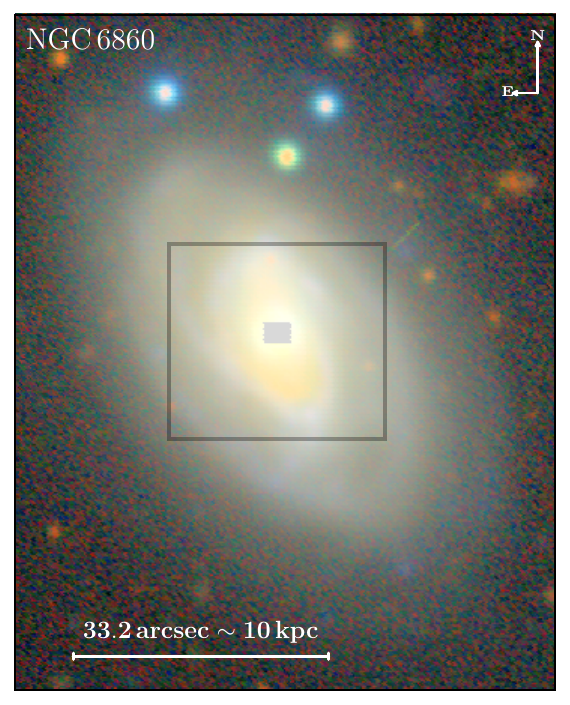}
    \includegraphics[width=0.285\textwidth]{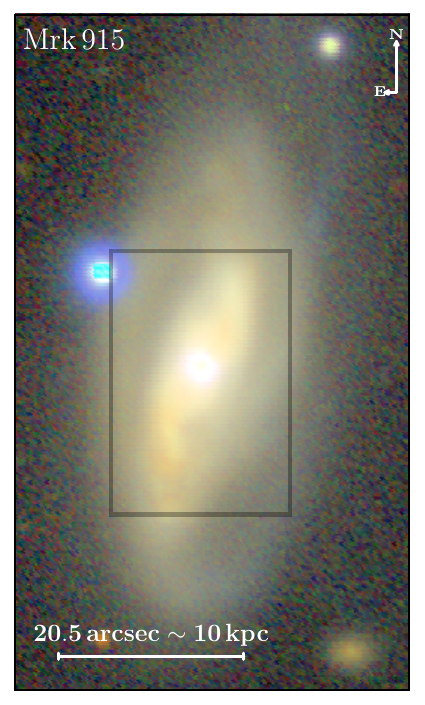}
    \includegraphics[width=0.323\textwidth]{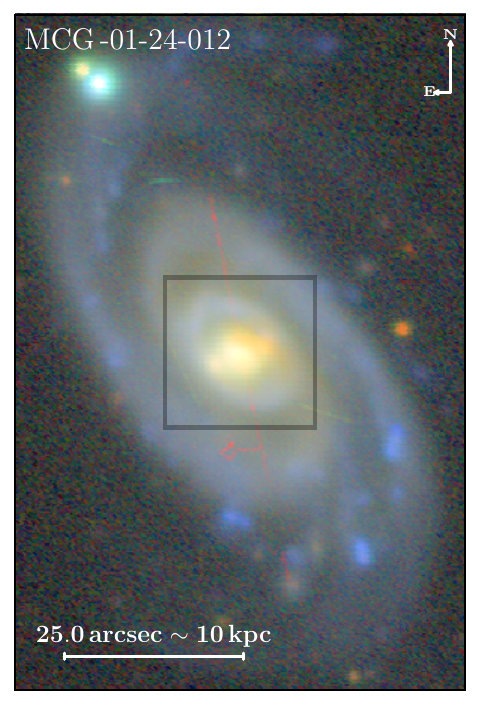}
\caption{Color-composite images of NGC\,6860, Mrk\,915 and MCG\,-01-24-012, generated with  g-r-i-z filters from the archival Dark Energy Camera images. North is up and East is left, as in all other figures in the paper. The rectangles correspond to the FoV of the maps in Fig.\,\ref{fig:upper}. 
The central gray rectangle on the left image masks an artifact present in the i and z-band images of NGC\,6860. In Appendix\,\ref{ap:rgb}, we describe how these figures were generated.}
\label{fig:fig0}
\end{figure*}

The three nearby Seyfert galaxies studied in this paper -- NGC\,6860, Mrk\,915 and MCG\,-01-24-012 - are part of an ALMA set of proposals (2012.1.00474.S, 2015.1.00086.S, 2018.1.00211.S) that aimed to analyze in detail the cold molecular gas in 13 local active galaxies, that we will call the ``original sample''. These objects were selected for having signs of disturbance in the ionized phase, including collimated outflows and/or the presence of nuclear spirals associated with inflows. The other 10 galaxies from this project have been studied in group \citep{ramakrishnan+19} or in individual studies from different authors \citep{finlez+18,slater+19,salvestrini+20,dallagnol+21,rosario+19,feruglio+20,shimizu+19}. 
Here, we complete the analysis by studying the three remaining sources together and comparing the results with the other objects from the sample. They are presented in the color-composite images in Fig.\,\ref{fig:fig0}, generated from archival DECam (Dark Energy Camera) images, obtained from DESI (Dark Energy Spectroscopic Instrument) Legacy Imaging Surveys (see Appendix\,\ref{ap:rgb}).

Table \ref{tab:sample} shows some basic properties of the sample. 
The redshifts (z) correspond to the galaxies' systemic velocities ($\vel_{\rm{sys}}$) and were obtained from kinematic 2D disk models fitted to data in the regions dominated by rotation \citepalias{dallagnol+25_paperII}. 
These redshifts were used to obtain the luminosity distances (\DL) and the angular scales, by assuming a $\mathrm{H_0=70\,km\,s^{-1}}$, $\mathrm{\Omega_M=0.3}$ and $\mathrm{\Omega_\Lambda=0.7}$ cosmology.

These Seyfert galaxies, with 0.014 < z < 0.025, have AGN total luminosities of 10$^{43.6}$\,$\lesssim$\,{\lagn}\,$\lesssim$10$^{44.8}$\,{\ergs} and supermassive black hole masses of 10$^{7.1}$\,$\lesssim$\,{\MBH}\,$\lesssim$10$^{8.4}$\,{\msun}, with their galaxy hosts having total stellar masses of 10$^{9.6}$\,$\lesssim$\,{\Mstar}\,$\lesssim$10$^{10.3}$\,{\msun} (all references on Table\,\ref{tab:sample}). 
They were classified as Seyfert type 1.5, 1.9 or 2, according to the strength of the broad line region component of the H$\beta$ line \citep{osterbrock77}.

\begin{figure*}
     \centering
    \begin{subfigure}[b]{1\textwidth}
        \centering
    	\includegraphics[width=1\linewidth]{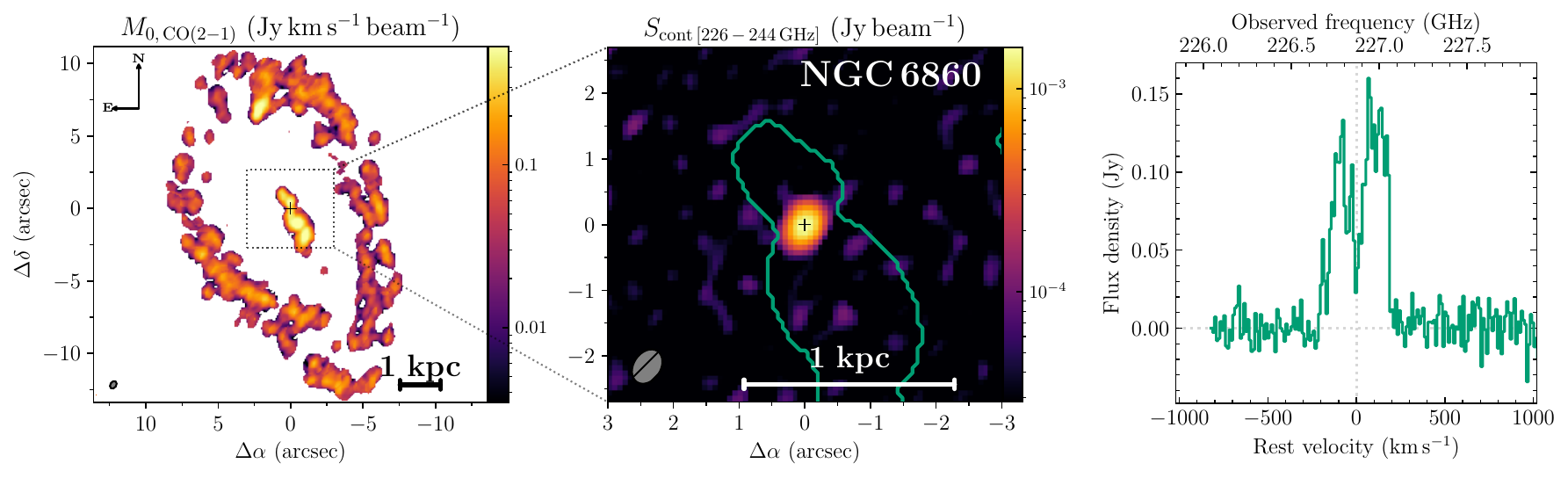}
    \end{subfigure}
    \begin{subfigure}[b]{1\textwidth}
        \centering
    	\includegraphics[width=1\linewidth]{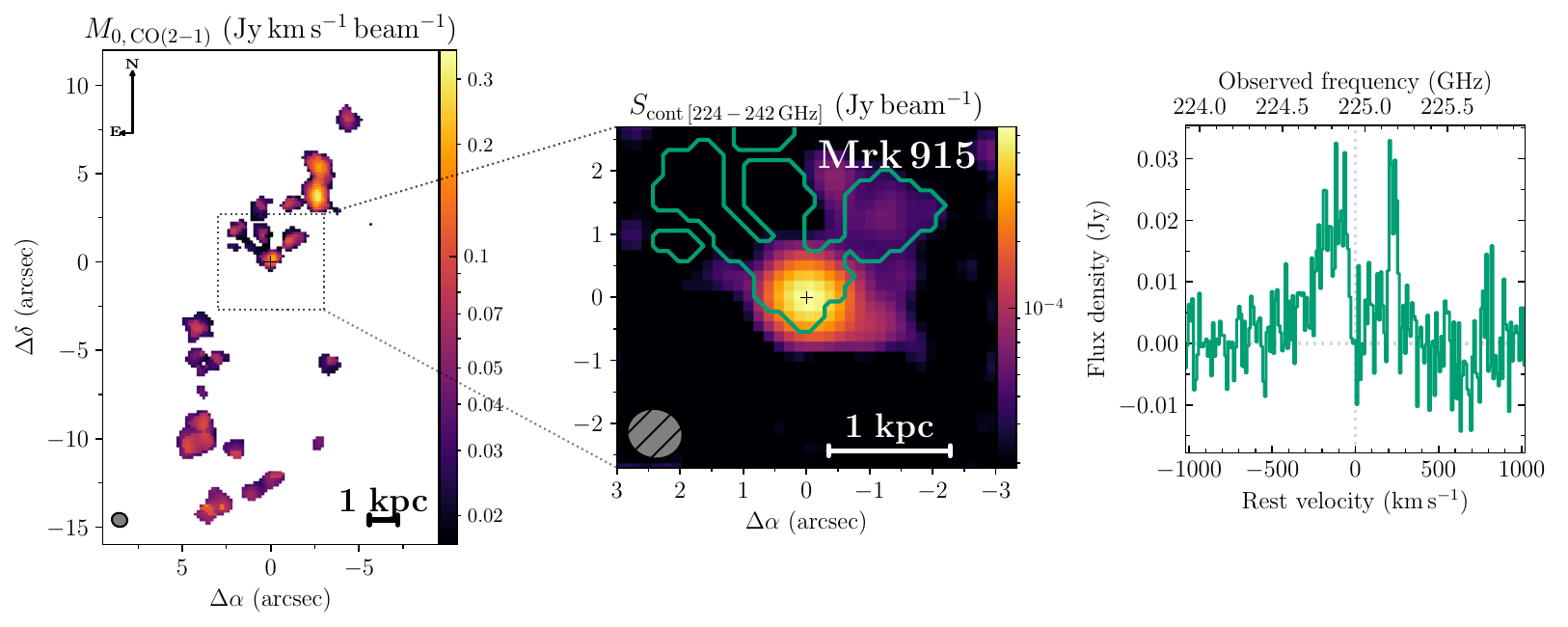}
    \end{subfigure}
    \begin{subfigure}[b]{1\textwidth}
        \centering
    	\includegraphics[width=1\linewidth]{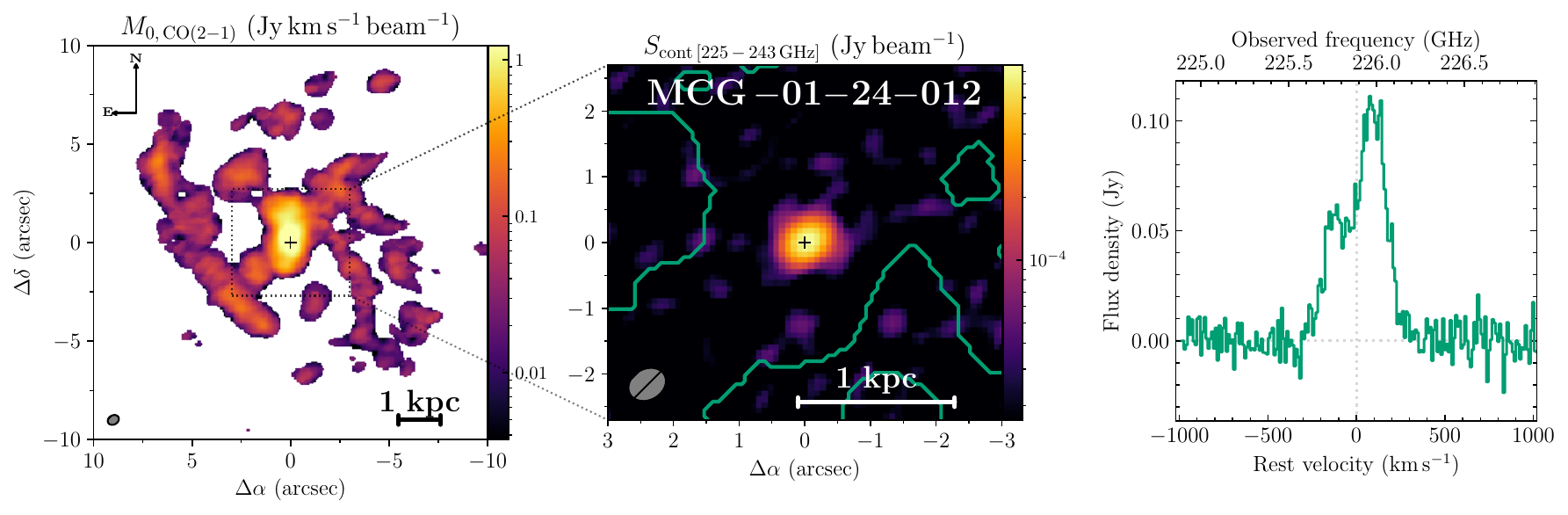}
    \end{subfigure}
    \caption{
        Overview of the final reduced ALMA data, showing one object per row. 
        Left: The CO(2-1) moment $M_0$, corresponding to the integrated flux. 
        Center: ALMA millimeter continuum flux density near the CO spectral region, with $M_0$ outline here with a bluish-green contour. 
        Right: CO(2-1) spectral profile, integrated over the whole FOV (only spaxels not masked on the $M_0$ map), showing the flux density vs. rest velocity (or observed/sky frequency, in the top). 
        The coordinates are relative to the galactic nucleus (black cross marks), assumed to be located at the ALMA millimeter continuum peak (see Table\,\ref{tab:sample}).  
        The gray ellipses represent the ALMA beam resolution of the observation. 
        }
        \label{fig:upper}
\end{figure*}

\section{Observations}\label{sec:obs}

\subsection{ALMA}\label{sec:obs-alma}
The three galaxies were observed with ALMA in Cycle 6 (ID: 2018.1.00211.S, PI: Ramakrishnan, V.). One of the spectral windows (SPW) was centered on the {\coII} emission line. In addition, three other SPW were centered on the nearby continuum. These were used to image the nuclear continuum and to subtract the underlying continuum from the line profiles. The total observing time on source for NGC\,6860, Mrk\,915 and MCG\,-01-24-012 were 33.9, 16.2 and 29.9\,min, respectively.
Table\,\ref{tab:obs_mass} shows other relevant information about the observations.

We used the pipeline scripts provided by ALMA archive to reprocess the archival data using the CASA software \citep{casa22}. Using this script, we performed typical reduction steps, including cross-calibration, flagging, and bandpass calibration. 
For the imaging, we used the tclean CASA's task with the Briggs weighting scheme \citep{briggs95},  which is controlled by the ``robustness'' parameter: a value of $-2$ is close to the uniform weighting, where the angular resolution is maximized at the expense of the sensitivity, with the opposite being true for a value of 2 (natural weighting).

For NGC\,6860 and MCG\-01-24-012, we used a ``robustness'' parameter of 0 and 0.5, respectively. 
To improve the signal-to-noise ratio (S/N) of the Mrk\,915, we used a ``robustness'' value of 2 to recover of the CO emission of Mrk\,915 over a larger region inside the field-of-view (FoV). 
With the same goal, we also applied a Gaussian ``uv-taper'' of 0.5{\arcsec}, to partially fill holes in the uv-space. 
As expected, this weighting scheme resulted in a poorer angular resolution in Mrk\,915 compared to the other two objects, as shown by the full width at half maximum (FWHM) of the beam in the table. 
The final spatial resolution is $\sim$\,0.5\,--\,0.8{\arcsec} ($\sim$\,150\,--\,400\,pc), from the mean FWHM$_{\rm{beam}}$. 
The angular sampling (spaxel sizes) of the cubes are 0.076{\arcsec}, 0.16{\arcsec} and 0.087{\arcsec} for NGC\,6860, Mrk\,915 and MCG\,-01-24-012, respectively.

To increase the S/N, the data cubes of the three objects were binned in pairs of two channels, with the resulting data cube having a channel width of $\Delta \vel$\,$\sim$\,10.2\,{\kms}.
Table\,\ref{tab:obs_mass} also shows the corresponding {\sigmarms} noise per channel at the center of the FoV. It corresponds to the standard deviation of the flux density in line-free regions prior to the primary beam correction (sensitivity over the FoV, a product from the data reduction). Throughout the paper, we used a 2D version \sigmarms(x,y) = \sigmarms\,$/$\,PB(x,y), where PB(x,y) is the primary beam map, which better replicates the increase in the noise towards the image edges.

An overall picture of the resulting reduced data is presented in Fig.\,\ref{fig:upper}. The maps of total CO(2-1) flux and the millimeter continuum flux density are shown in the left and middle columns, while the integrated CO(2-1) profile is shown in the right column. The continuum maps display an overall point-like spatial distribution, with only Mrk\,915 showing a small elongation to the North-West.

The right ascensions (RA) and declinations (DEC) in Table\,\ref{tab:sample}
correspond to the spatial coordinates of the peak of the ALMA  1.3\,mm (230\,GHz) continuum (middle column of Fig.\,\ref{fig:upper}, and marked as a cross in the other maps). 
We use this peak as the location of the AGN based on the correlation between 1\,mm luminosity ($L_{\mathrm{1\,mm}}$) with the 2\,–\,10\,keV X-ray luminosity ($L_{\mathrm{2{-}10\,keV}}$) and the black hole mass ({\MBH}) in a plane \citep{ruffa+24}). Comparing only with the X-ray continuum, there is a tighter correlation with $L_{\mathrm{1\,mm}}$ and $L_{\mathrm{14{-}150\,keV}}$ then $L_{\mathrm{2{-}10\,keV}}$ \citep{kawamuro+22}. 
Nonetheless, the precise origin of the 1\,mm emission is still uncertain and possibly not unique \citep{salvestrini+20,kawamuro+23,shablovinskaya+24}, with some of the proposed explanations for the nuclear emission being small-scale jets and advection-dominated accretion flows (ADAFs) \citep{ruffa+24}. 
At the ALMA and {\hst} spatial resolutions used here, the positions are expected to be coincident.

\begin{table*}
\centering
\caption{
Information about ALMA observations, data reduction and the total molecular mass.
(1) Galaxy name;
(2) ALMA synthesized beam (FWHM of the minor $\times$ major axis);  
(3) Mean spatial resolution (from the mean of FWHM$_{\rm{beam}}$); 
(4) Noise at the FoV center, for a $\Delta \vel$\,$\sim$\,10.2\,{\kms} channel width; 
(5) ``robustness'' parameter of the Briggs weighting; 
(6) Total CO(2-1) flux; 
(7) Total CO(1-0) luminosity (for $r_{21}$\,=\,0.8\,--\,1.2); 
(8) Total cold molecular gas mass ({\alphaCO}\,=\,0.8\,--\,4.3\,{\msunKkmspc}).
}
\centering
\begin{tabular}{lcccccccr}
\hline
            Name &             FWHM$_{\rm{beam}}$ &     resolution & \sigmarms &         robustness & $S_{\nu} \Delta \vel_{\rm{CO(2-1),tot}}$ & $L'_{\rm{CO(1-0),tot}}$ & $M_{\rm{mol,tot}}$  \\
                 & $\mathrm{arcsec^{2}}$ &        pc & $\mathrm{mJy\,beam^{-1}}$ &                    & {\jykms} &       $10^7$\,{\Kkmspc} &    $10^7$\,{\msun}  \\
       (1) &  (2) &  (3) &  (4) &  (5) &  (6) &  (7) &  (8)   \\

\hline
       NGC\,6860 &  0.41$\times0.56$ &  150 & 0.69 & 0       & $47.18\,{\pm}\,0.05$ &         $9.7$\,--\,$15$ &      $8$\,--\,$60$  \\
        Mrk\,915 &  0.78$\times0.88$ &   400        & 0.53 & 2$^{a}$ & $4.99\,{\pm}\,0.02$ &        $2.8$\,--\,$4.2$ &      $2$\,--\,$20$  \\ 
 MCG\,-01-24-012 &   0.47$\times0.6$ &   210         & 0.44 & 0.5     & $31.22\,{\pm}\,0.03$ &          $11$\,--\,$17$ &      $9$\,--\,$70$  \\
\hline
\end{tabular}
{
\\
\raggedright
$^{a}$: with an additional 0.5\,{\arcsec} ``uv-taper''\par
}
\label{tab:obs_mass}
\end{table*}

\subsection{Archival data}\label{sec:obs-other}\label{sec:data_literature}

For the ionized gas distribution, we used {\hst} FR533N narrow-band images \citep[ID: 8598,][]{schmitt+03}, which covers the {\oiii}$\lambda\lambda$4959,5007 emission lines region. 
To isolate the emission from the {\oiii} lines, F547M continuum images (from the same observation proposal) were subtracted from the narrow-band {\oiii} ones. 
As described in the Appendix\,\ref{ap:oiii-continuum}, the continuum image is multiplied by a constant factor ($K_c$) before being subtracted from the narrow-band {\oiii} image. This factor is obtained interactively, by selecting the larger $K_c$ value that does not introduce negative regions (``depressions'') in the subtracted image. 
However, we note that the F547M filter has a passband that partially covers the {\oiii} line profiles, which may induce over-subtraction of the line, in some regions, particularly for NGC\,6860 (see the Appendix\,\ref{ap:oiii-continuum}). The sky background of the resulting continuum-subtracted {\oiii} images was subtracted, after fitting a 2D plane in regions dominated by noise. 

Visual inspection of the {\hst} images indicated that they were spatially misaligned relative to the remaining images. 
Therefore, we modified their CRPIX/CRVAL header key values, such that the {\hst} continuum peak matched the ALMA millimeter continuum peak. The CRPIX/CRVAL values for {\hst} {\oiii} images were set equal to those of their respective F547M continuum images, which were aligned before. Comparing the resulting position of stars in the field in the {\hst} and DECam images, we expect a maximum error in astrometry of $\sim$\,0.1\arcsec.

All {\hst} images were collected from the Mikulski Archive for Space Telescopes (MAST), selecting pipeline drizzled products. The re-processing discussed above was applied to these images.  The DECam images are from the Data Release 10 (DR10) of the DESI Legacy Imaging Surveys.

For MCG\,-01-24-012 and Mrk\,915, we also collected \textit{Very Large Array} ({\vla}) archival 8.46\,GHz (3.54\,cm) radio images \citep[proposal ID: AA226]{schmitt+01}.
The {\vla} images were downloaded from the NASA Extragalactic Database, without further modifications. 

\section{Analysis}\label{sec:analysis}

\subsection{Structure maps}\label{sec:structure_maps}
\begin{figure*}
	\includegraphics[width=1\linewidth]{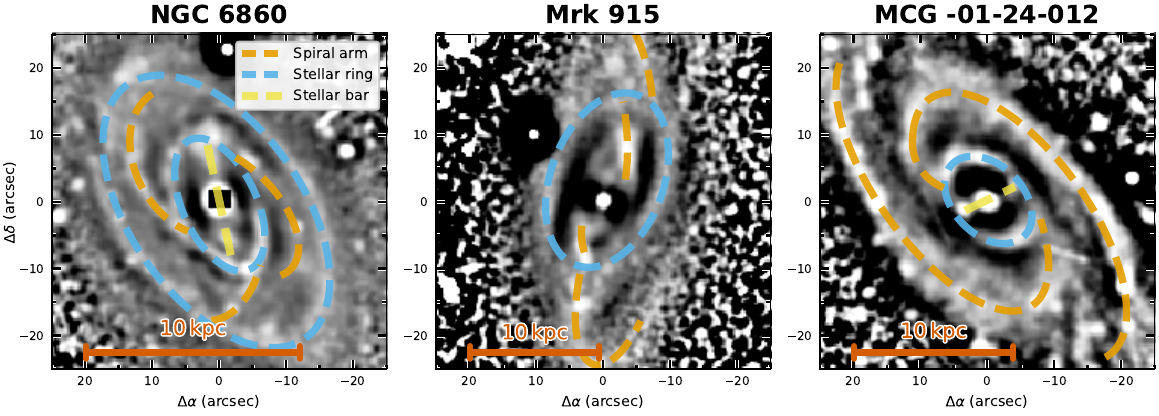}
    \caption{Structure maps, with dashed lines/curves highlighting spiral arms (in orange), stellar rings (sky-blue), and bars (yellow). 
    The lines were drawn by visual inspection of the top structure maps, obtained for the archival DECam i-band images, where structures with spatial scales of $\sim$500\,pc are enhanced.  
    }
    \label{fig:structure_maps}
\end{figure*}

To highlight morphological stellar features in the host galaxies, we constructed structure maps using images of the sources, as shown in Fig.\,\ref{fig:structure_maps}. With such a technique, structures with scales of the order of the Point Spread Function ($\sim$\,500\,pc) are enhanced \citep{pogge_martini02}. 

Following \citet{simoes_lopes+07}, these structure maps ($S$) were obtained using \citep{pogge_martini02}: 
\begin{equation}
    S = \left[\frac{I}{I\otimes P}\right] \otimes P^t,
\end{equation}
where I is the original image, and P is the Point Spread Function. 
To obtain the PSF, we modeled the flux distributions of field stars with Moffat profiles \citep[e.g.][]{trujillo+01}.
The fitting and convolution procedures were performed using {\sc astropy.modeling} and {\sc astropy.convolve}. 

Dashed lines in Fig.\,\ref{fig:structure_maps} mark the location of spiral arms, rings and bars, all identified by visual inspection.
We obtained the structure maps using DECam i-band images (covering 710\,--\,860\,$\rm{nm}$) because we wanted to highlight stellar bars, and the light from such structures is dominated by the contribution from old stars, whose peak of emission is seen at longer wavelengths. 

\subsection{Molecular gas mass}\label{sec:mass}

The total velocity-integrated flux ($S_{\nu} \Delta \vel_{\rm{CO(2-1)}}$) and the derived CO(2-1) luminosity ($L'_{\rm{CO(2-1)}}$) and molecular mass ($M_{\rm{mol}}$) are presented in Table\,\ref{tab:obs_mass}. 
The CO flux -- in each spaxel -- was derived from the Gaussian profiles fitted to the CO profile, with a maximum of two components being needed to reproduce each profile \citepalias{dallagnol+25_paperII}.
The corresponding uncertainties are obtained following \citet{lenzs_ayres92} and are propagated when integrated over a region and/or including more than one component. 
The corresponding luminosities ($L'_{\rm{CO(2-1)}}$), in units of {\Kkmspc}, are obtained using the Equation 1 from \citet{solomon+97}. We converted the resulting values to CO(1-0) luminosities by using the ratio of $r_{21}$\,=\,$L'_{\rm{CO(2-1)}}/L'_{\rm{CO(1-0)}}$\,=\,0.8\,--\,1.2, as  observed in nearby galaxies \citep{braine_combes92}.

The molecular mass can then be calculated using the conversion factor
{\alphaCO}\,=\,$M_{\rm{mol}}/L'_{\rm{CO(1-0)}}$. 
To cover different cloud physical conditions, we assumed a large range of values: {\alphaCO}\,=\,$M_{\rm{H_2}}/L'_{\rm{CO(1-0)}}$\,=\,0.8\,--\,4.3\,{\msunKkmspc}. The upper limit value in the above {\alphaCO} corresponds to the average in the Milk Way inner disk, while the lower limit is the average in Ultra Luminous Infrared Galaxies \citep{bolatto+13}.

In addition to the large uncertainties in the CO-to-H$_2$ conversion, we note that the total CO(2-1) luminosities, used in the calculation, may be systematically underestimated.
This happens because interferometric observations filter out large-scale emission, 
causing low-surface brightness CO(2-1) emission to be missed. This issue is sometimes referred to as the ``missing flux problem''. As a consequence, our total CO fluxes and $M_{\rm{mol,tot}}$ measurements can be considered lower limit values.

To illustrate this, we can compare our total flux measurements with values obtained using single-dish telescopes. For NGC\,6860, \citet{strong+04} obtained CO(2-1) observations with the Swedish–ESO Submillimetre Telescope (SEST, with FWHM$_{\rm{beam}}$ of 22{\arcsec}), resulting in $L'_{\rm{CO(2-1)}}$\,=\,23$\times10^{7}$\,{\Kkmspc} (value corrected to consider the same cosmology parameters assume here). 
This value is a factor of $\sim$\,2 larger than our measurement of $L'_{\rm{CO(2-1)}}$\,$\sim$\,11.8$\times10^{7}$\,{\Kkmspc}. 
Similarly, \citet{koss+21} presented observations of our objects from the Atacama Pathfinder Experiment telescope (APEX, FWHM$_{\rm{beam}}$ of 27{\arcsec}), with CO(2-1) luminosity values that are larger than our measurements by a factor of $\sim$\,3 for NGC\,6860 and MCG\,-01-24-012, and $\sim$\,6 for Mrk\,915. 
In addition, CO emission from regions beyond the FoV of our ALMA observations may also contribute to the single dish measurements. 
For comparison, the FWHM$_{\rm{beam}}$ of these SEST and APEX observations are, respectively, $\sim$\,1.3 and 1.7 times larger than our observations FOV ({$\sim$}\,17{\arcsec} radius, at one-fourth of the sensitivity of the FoV center).

\section{Results: CO vs. {\oiii} relation}\label{sec:results}

\begin{figure*}
	\includegraphics[width=1\linewidth]{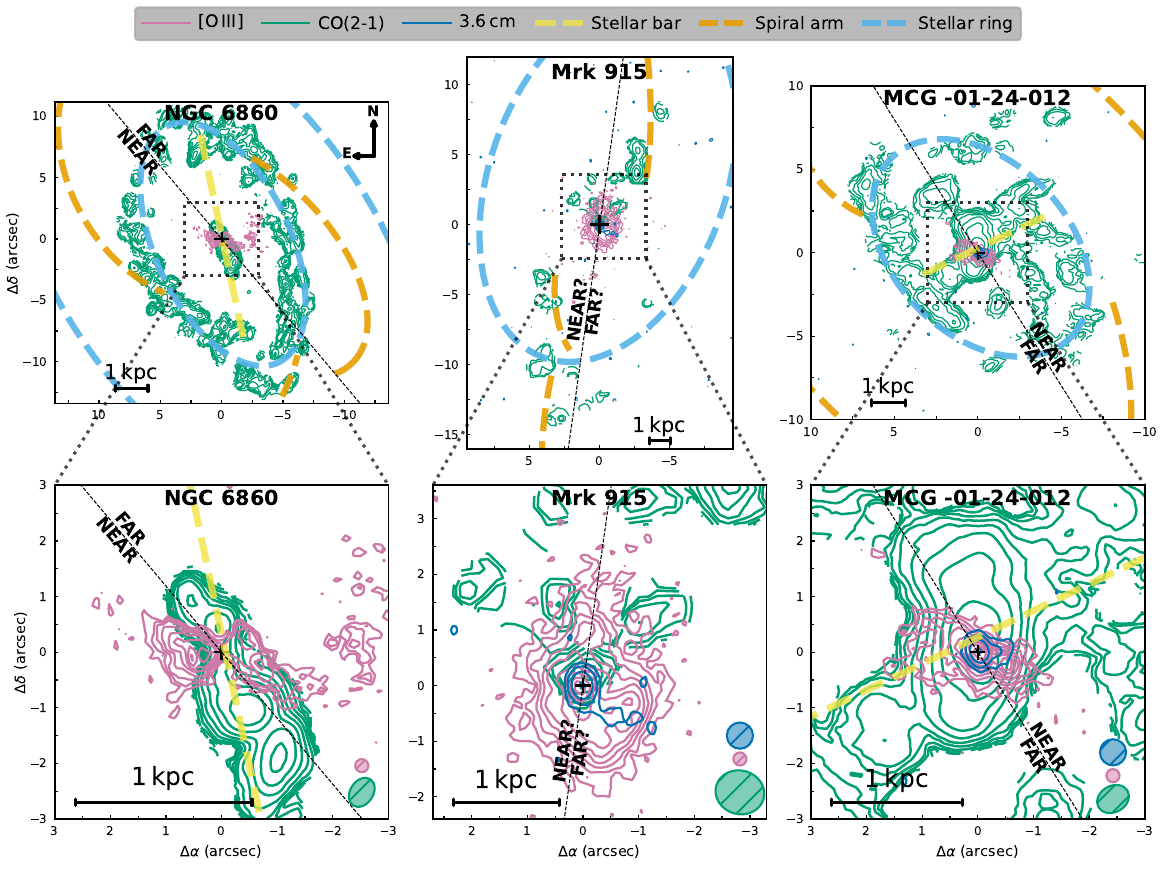}
    \caption{Contours showing the flux distribution from the CO(2-1) cold molecular gas ($M_0$, in bluish-green), {\oiii} ionized gas (reddish-purple)  and the 3.6\,cm radio (blue), with a zoom-in of the nuclear 3\arcsec$\times$3{\arcsec} region (bottom pannels). 
    The yellow, orange and sky-blue  dashed lines correspond to stellar bar, spiral arm and stellar ring structures, respectively (as identified in Fig.\,\ref{fig:oiii_co_grid}). 
    Colored ellipses represent the beam/PSF resolutions of each contour image, using the same color of the tracer. 
    Contours are evenly spaced logarithmically. 
    There are: 
    10 CO(2-1) contour levels in the ranges (0.508, 47.8), (0.620, 28.0) and (0.359 93.5)\,{\jykms}, for NGC\,6860, Mrk\,915 and MCG\,-01-24-012;
    10 {\oiii} levels in the ranges (10$^{-18.3}$, 10$^{-16.4}$), (10$^{-18.2}$, 10$^{-15.7}$) and (10$^{-18.3}$, 10$^{-16.2}$)\,{\ergscmAA}, in the same order; 
    4 radio levels in the ranges (10.2, 1202) and (10.35, 713)\,mJy, for Mrk\,915 and MCG\,-01-24-012.
    }
    \label{fig:oiii_co_grid}
\end{figure*}

In the central kiloparsec, the flux distributions of the {\oiii} and the CO(2-1) are spatially anti-correlated in our sample.  
This is shown by the contours comparing the CO(2-1) and {\oiii} flux distributions in Fig.\,\ref{fig:oiii_co_grid}. 
Also displayed is the 3.6\,cm radio emission available for Mrk\,915 and MCG\,-01-24-012.

In NGC\,6860, the CO(2-1) emission is observed along the stellar ring and the bar (inside a $r$\,$\sim$\,4\,kpc projected radius from the nucleus), corresponding to total cold molecular gas mass of $M_{\rm{mol,tot}}$\,$\sim$\,8$\,$--$\,$60{\,$\times\,10^{7}\,$\msun}.
We notice that, close to the nucleus ($r$\,$\sim$\,500\,pc), the CO emission presents an hourglass shape: the projected emission shows a narrowing width as it approaches the nucleus, decreasing from $\sim$\,400\,--500\,pc to $\sim$\,250\,pc width (lower left map of Fig.\,\ref{fig:oiii_co_grid}).
Interestingly, the {\oiii} bipolar emission -- extending along the East-West direction -- seems to fill the region where the CO flux distribution is narrower, with the {\oiii} contours forming ``bowl-shape'' at the base of the emission. 

In Mrk\,915, the cold molecular gas is observed along the two central spiral arms, totalizing $M_{\rm{mol,tot}}$\,$\sim$\,2$\,$--$\,$20{\,$\times\,10^{7}\,$\msun}. Unlike the other two Seyferts, the {\oiii} ionized gas in Mrk\,915 does not present a bipolar morphology, in the sense of the {\oiii} being elongated in opposite/aligned directions (lower middle panel in Fig.\,\ref{fig:oiii_co_grid}). Instead, it is spatially elongated along two directions:  North (PA\,$\sim$\,5{\degree}) and South-West of the nucleus (PA\,$\sim$210\,--\,235{\degree}). Also extended to the South-West of the nucleus, there is a weak 3.6\,cm radio source that could indicate the projected direction of a young radio-jet axis, although further observations are needed to better resolve and confirm the jet. Additionally, to verify a young-source origin for the emission, it is required to calculate the radio spectral index, which depends on the availability of multi-band radio observations \citep[e.g.][]{kukreti+23}. 
Nonetheless, except in the nucleus, we still observe a spatial anti-correlation between CO and {\oiii} flux distributions.

In MCG\,-01-24-012, most of $M_{\rm{mol,tot}}$\,$\sim$\,9$\,$--$\,$70{\,$\times\,10^{7}\,$\msun} cold molecular mass, as traced by CO emission, is detected in the stellar ring and the area within it. Within this region, the CO flux distribution is not only observed/concentrated near the bar (like in NGC\,6860), being more spread. 
Within the inner 2.5{\arcsec}$\sim$\,1\,kpc radius, the strongest CO emission is distributed along the North-South direction, with a small twist to the North-East at larger radii (right panel in Fig.\,\ref{fig:oiii_co_grid}). 
This distribution is approximately perpendicular to that of the {\oiii} bipolar emission, that has PA\,$\sim$\,75{\degree} \citep{schmitt+03}.
We also note that the 3.6\,cm radio emission has a small elongation to the West \citep[PA\,$\sim$\,89{\degree}, ][]{schmitt+01}, approximately aligned with the {\oiii} ionization axis.

\section{Discussions}
\label{sec:Discussion}

A strong spatial anti-correlation between the cold molecular versus the ionized gas and/or the radio jet was already observed in other active galaxies \citep[e.g.][]{garcia-bernete+21,audibert+23,alonso-herrero+18}. We found the same trend for the three sources studies here: NGC\,6860, Mrk\,915 and MCG\,-01-24-012. What about the remaining 10 objects in our original sample (13 sources in total)?

From these 10 sources, 6 have individual studies comparing the CO vs. jets/ionized gas emissions. 
And 5 of them -- all showing extended bipolar ionized cones/jets -- follow the above anti-correlation trend: NGC\,3281 \citep{dallagnol+23}, NGC\,3393 \citep{finlez+18}, ESO\,428-G14 \citep{feruglio+20}, NGC\,2110 \citep{rosario+19}, NGC\,5728 \citep{shimizu+19}. 
In NGC\,1566, there is a CO deficit in the nucleus \citep{slater+19}, where ionized gas emission is unresolved under $\sim$\,50\,pc, \citep{dasilva+17}. For the remaining 4 objects, we performed a visual inspection using CO and ionized gas images from different studies.
We identified a CO gap \citep{ramakrishnan+19,salvestrini+20} associated with the bipolar ionization cone in NGC\,1386 \citep[$\sim$\,100\,pc to the North of the nucleus]{rodriguez-ardila+17}, NGC\,7213 \citep[$\sim$\,300\,pc to the South-West]{schnorr-muller+14} and NGC\,3081 \citep[South-West to North-East direction, under a $\sim$\,600\,pc radius]{schnorr-muller+16}.
In NGC\,1667, a local CO deficit is visible \citep{ramakrishnan+19} where the ionized gas emission is slightly elongated ($\sim$\,400\,pc to the East) and where it is more disturbed \citep[South-East to North-West direction]{schnorr-muller+17}. Overall, we found some degree of spatial anti-correlation between CO and ionized gas/jets in 12 out of 13 objects from the original sample. 
What is the cause behind it?

To understand this, we can first focus on the three sources presented here. A possible scenario is that part of the molecular gas is being dissociated by AGN radiation/winds/jets. 
This assumes that the {\oiii} extended emission traces the AGN ionization axis, which is corroborated by the bipolar {\oiii} flux distribution in NGC\,6860 and MCG\,-01-24-012. Also, Mrk\,915 and MCG\,-01-24-012 show weak off-nuclear 3.6\,cm radio emission, which extends in one of the directions where the {\oiii} emission is elongated. 

In favor of this, in NGC 2110, \citet{kawamuro+20} also observed an anti-correlation between the flux distribution of the CO and the Fe-K$\alpha$\,6.4\,keV fluorescent line, which needs ionization from hard X-ray photons to be produced. This reinforces the scenario that the CO or/and H$_2$ may be depleted due to the high-energy photons from the AGN. A similar conclusion may be drawn for NGC\,5728, where the X-ray emission containing highly ionized lines is observed in the region devoid of CO gas \citep{trindade_falcao+24}. However, the presence of cold scattering material may indicate that part of the molecular gas could still be present along the ionization axis (see below).

We emphasize that an observed spatial anti-correlation between CO emission and ionized gas is not definitive evidence that molecular gas is being destroyed. For example, the AGN radiation could be ionizing the gas located in (or the ionized outflow could be expanding to) regions already devoid of cold molecules. Besides that, even if we follow the hypothesis that CO emission gaps are a consequence of CO molecules being partially destroyed, 
H$_2$ molecules might still survive in these depleted regions. 

For example, in NGC\,2110,  emission from the hot H$_2$ 1-0 S(1) line\footnote{Near-infrared rovibrational transition,  $T_{\rm{gas}}$\,$\sim$\,10$^3$\,K} is detected in a CO(2-1) lacuna \citep{rosario+19}. 
A similar scenario is observed in the Seyfert galaxies NGC\,5643 \citep{alonso-herrero+18,davies+14} and ESO 428-G14 \citep{feruglio+20}, both on CO(2-1) and H$_2$ 1-0 S(1) transitions. 
Another example is NGC\,5728, with emission from the warm H$_2$ 0-0 S(3) line\footnote{Mid-infrared pure rotational transition, $T_{\rm{gas}}$\,$\sim$\,10$^2$\,--\,10$^3$\,K} being observed throughout the ionization cone, where there is a CO deficit \citep{davies+24}. 
Other warm H$_2$ 0-0 S(1), S(5) \citep{davies+24} and hot  H$_2$ 1-0 S(1) lines \citep{shimizu+19} are also detected in the gap of the CO in NCG\,5728.  
In MCG\,-05-23-16, warm H$_2$ gas (from multiple transitions) is detected in the CO(2-1) cavities, in a region where H$_2$ displays higher velocity dispersion, likely driven by recent star formation in this case \cite{esperanza+25}.

After comparing mass surface densities on 50 and 200\,pc scales in Seyfert galaxies, \citet{garcia-burillo+24} found a decrease in the concentration on both the cold and hot molecular gas phases for X-ray luminosities above L$_{\rm{2-10\,KeV}}$\,$\sim$\,10$^{41.5}$\,{\ergs}. 
They interpret that the AGN wind/radiation pressure or the jet has pushed out the H$_2$ gas since the most extreme nuclear-scale cold molecular gas deficits are observed in galaxies with the strongest CO outflows. 
This is in agreement with our results, since the X-ray luminosities of our objects are a factor of $\sim$\,10 above this threshold \citep{duras+20}. Nonetheless, their deficit in the cold molecular gas is not fully compensated (filled) by the hot molecular phase emission.  

We note that in high spatial resolution (50\,--\,100\, pc scales) observations of star-forming regions, the CO vs. {\ha} distribution shows that the cold molecular gas is also anti-correlated in areas with recent star-formation \citep{leroy+21, kreckel+18}, although having some overlap in denser areas, as in spiral arms \citep{larson+23}.
The ionizing radiation from the newly born stars is likely depleting the nearby CO molecular clouds inside {\hii} regions (traced by {\ha}). This is similar to what is observed in our sources, but for a weaker ionization field. 

As a final remark, we point out that the maximum recoverable scale of emission in our ALMA observations is $\sim$\,6{\arcsec} ($\sim$\,2\,--\,3\,kpc). Emission from larger scales will therefore not be detected in our data. This could be why we measure  $\sim$\,2\,--\,6 times lower total CO flux compared to the single-dish observations (see Section\,\ref{sec:mass}), which do not suffer from the missing flux problem. However, this is unlikely to be the reason behind the observed CO gaps, since they are identified on smaller scales ($\lesssim$\,1\,kpc, see Fig.\,\ref{fig:oiii_co_grid}). Any CO emission present at these scales would have been detected with the sensitivity of the ALMA data. We therefore propose that these CO gaps show a real inhomogeneity in the CO distribution.

\section{Conclusions}\label{sec:conclusions}

We have compared the CO cold molecular gas distributions -- as traced by CO(2-1) ALMA observations -- with that of the {\oiii} ionized gas distributions -- from {\hst} images -- of three nearby Seyfert galaxies and compared with results for the original sample of 13 active galaxies.
The main conclusions are:

\begin{itemize}
    \item There is a spatial anti-correlation between the cold molecular gas and ionized gas distributions in the three galaxies studied, observed within the inner kpc. 
    Evidence of the same trend is presented in other studies in the literature, including 12 out of 13 nearby AGN sources from our original sample.
    \item The anti-correlation can be attributed -- in part -- to a depletion of the cold molecular gas due to the AGN radiation, winds and/or jets. Warmer phases of the molecular gas (e.g. infrared H$_2$ lines) might fill part of the CO gap region.
\end{itemize}

Assuming that the observed CO deficit is due to an interaction between the AGN released energy and the ISM, one can argue that this is a type of AGN negative feedback not accounted for when measuring the impact using quantities like the mass outflow rate. 
In our following \citetalias{dallagnol+25_paperII}, we present measurements of this impact obtained 
from the analysis of the CO(2-1) kinematics. However, to fully discriminate how much of the molecular gas is completely depleted, measurements of the distribution and total mass of the molecular gas at different temperatures are needed.

\begin{acknowledgements}

This study was financed in part by the Coordena{\c c}{\~a}o de Aperfei{\c c}oamento de Pessoal de N\'ivel Superior (CAPES-Brasil, 88887.478902/2020-00, 88887.985730/2024-00). 
RAR acknowledges the support from Conselho Nacional de Desenvolvimento Cient\'ifico e Tecnol\'ogico (CNPq; Proj. 303450/2022-3, 403398/2023-1, \& 441722/2023-7), Funda\c c{\~a}o de Amparo \`a pesquisa do Estado do Rio Grande do Sul (FAPERGS; Proj. 21/2551-0002018-0), and CAPES (Proj. 88887.894973/2023-00).\\

This paper makes use of the following ALMA data: ADS/JAO.ALMA\#2018.1.00211.S, ADS/JAO.ALMA\#2015.1.00086.S, ADS/JAO.ALMA\#2012.1.00474.S. ALMA is a partnership of ESO (representing its member states), NSF (USA) and NINS (Japan), together with NRC (Canada), MOST and ASIAA (Taiwan), and KASI (Republic of Korea), in cooperation with the Republic of Chile. The Joint ALMA Observatory is operated by ESO, AUI/NRAO and NAOJ.\\

The Legacy Surveys consist of three individual and complementary projects: the Dark Energy Camera Legacy Survey (DECaLS; Proposal ID \#2014B-0404; PIs: David Schlegel and Arjun Dey), the Beijing-Arizona Sky Survey (BASS; NOAO Prop. ID \#2015A-0801; PIs: Zhou Xu and Xiaohui Fan), and the Mayall z-band Legacy Survey (MzLS; Prop. ID \#2016A-0453; PI: Arjun Dey). DECaLS, BASS and MzLS together include data obtained, respectively, at the Blanco telescope, Cerro Tololo Inter-American Observatory, NSF’s NOIRLab; the Bok telescope, Steward Observatory, University of Arizona; and the Mayall telescope, Kitt Peak National Observatory, NOIRLab. Pipeline processing and analyses of the data were supported by NOIRLab and the Lawrence Berkeley National Laboratory (LBNL). The Legacy Surveys project is honored to be permitted to conduct astronomical research on Iolkam Du’ag (Kitt Peak), a mountain with particular significance to the Tohono O’odham Nation.\\

This research has made use of the NASA/IPAC Extragalactic Database, which is funded by the National Aeronautics and Space Administration and operated by the California Institute of Technology.\\

This work used the color scheme from \citet{wong+11} in some of the plots to minimize confusion for colorblind viewers. 

\end{acknowledgements}



\bibpunct{(}{)}{;}{a}{}{,}
\bibliographystyle{aa_url}
\bibliography{bib}


\begin{appendix}

\section{RGB color-composite image}\label{ap:rgb}

To emphasize stellar structures in the central regions of our objects, we generated the griz color-composite images using archival DECam images from DESI. We applied the {\sc get\_rgb} function from the DESI Imaging Legacy Surveys viewer pipeline\footnote{\url{github.com/legacysurvey/imagine}}. This function implements an algorithm based on \citet{lupton+04}, 
with the RGB components obtained from a scaling of the griz-band filters images: 
R\,=\,(3\,$\cdot$\,i\,+\,2.2\,$\cdot$\,z), G\,=\,(3.4\,$\cdot$\,r) and B\,=\,(6\,$\cdot$\,g).
To highlight features in the central regions, we ran the original code, changing the default asinh softening value ($Q$) from 20 to 50 (see the result in Fig.\,\ref{fig:fig0}).

\section{{\oiii} continuum subtraction in NGC\,6860}\label{ap:oiii-continuum}

\begin{figure}
	\includegraphics[width=1\linewidth]{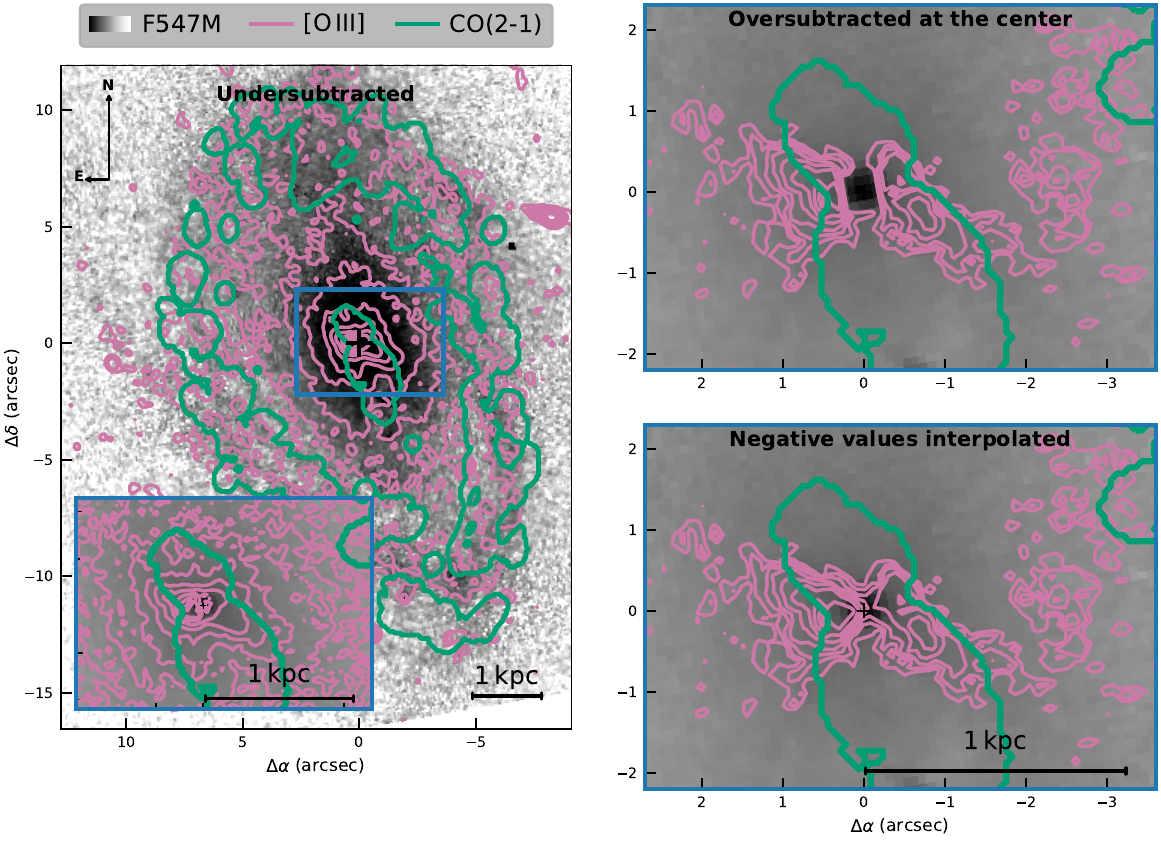}
    \caption{
    The effect on the {\oiii} flux distribution (reddish-puple contours) when the {\hst} F547M optical continuum image (in grayscale) is multiplied by different factors before the subtraction from the FR533N image (narrow band filter covering {\oiii}). 
    In the left, the continuum F547M was undersubtracted from {\oiii} -- using a lower multiplicative factor -- to avoid negative values are the center, with the inset showing a zoom-in in the central region. Notice that the smoothed {\oiii} contours follow the extended continuum distribution in this case. 
    On the right, the continuum was subtracted using a larger factor that resulted in negative values at the center (upper map). For the final image, the {\oiii} image was interpolated to eliminate the negative values (lower map). To show that the spatial anti-correlation between CO and {\oiii} is visible in both maps, we added a single blueish-green contour, delineating the CO(2-1) flux distribution. 
    In all zoom-in images, the F547M grayscale range is different maps to highlight the peak in the F547M image.} 
    \label{fig:oIII-co_ngc6860}
\end{figure}

The {\oiii} maps were obtained by subtracting the FR533N narrow-band image ({$\oiii$} + continuum) from the medium-band F547M image (nearby continuum) multiplied by a $K_c$ factor: {\oiii} = (FR533N)\,$-$\,$K_c\,\cdot$(\,F547M). This factor is interactively chosen, by maximizing the removal of the continuum contribution without introducing extended regions with negative values (continuum oversubtraction). However, this method introduces negative values in the central pixels of NGC\,6860 (upper right image of Fig.\,\ref{fig:oIII-co_ngc6860}).
The problem happens because a small part of the F547M band covers the {\oiii}$\lambda$5007 lines when its flux is stronger or its profile width is wider, resulting in an overestimation of the continuum at the nucleus. 

For the final {\oiii} image, the optical continuum was better subtracted using $K_c$\,=\,1.14. To remove the resulting negative values in the center, we interpolated the flux in this region (lower right image of Fig.\,\ref{fig:oIII-co_ngc6860}). The resulting {\oiii} contours are similar to the original data published in \citet{schmitt+03}. 

The left image of Fig.\,\ref{fig:oIII-co_ngc6860} shows the result of choosing a $K_c$ factor that avoids negative flux values at the nucleus. The {\oiii} would be more extended following the continuum, up to the stellar ring. The {\oiii} observations from \citet{lipari+93} show that there is {\oiii} emission in the stellar ring, but their total exposure time is $\sim$\,2 times longer.
Nonetheless, the nuclear {\oiii} morphology is consistent in the three cases of Fig.\,\ref{fig:oIII-co_ngc6860}).

\end{appendix}

\end{document}